\documentclass[twocolumn,showpacs,prc,superscriptaddress,amsmath,amssymb]{revtex4-1}

\usepackage{multirow}
\usepackage{graphicx}
\usepackage{dcolumn}
\usepackage{bm}
\usepackage{soul}
\usepackage{color}
\usepackage{array}

\begin{document}

\preprint{APS/123-QED}

\title{Three-body correlations in the ground-state decay of $^{26}$O}

\author{Z. Kohley}
\affiliation{National Superconducting Cyclotron Laboratory, Michigan State University, East Lansing, Michigan 48824, USA}
\affiliation{Department of Chemistry, Michigan State University, East Lansing, Michigan 48824, USA}
\author{T.~Baumann}
\affiliation{National Superconducting Cyclotron Laboratory, Michigan State University, East Lansing, Michigan 48824, USA}
\author{G.~Christian}
\altaffiliation[Present address: ]{TRIUMF, Vancouver, British Columbia V6T 2A3, Canada}
\affiliation{National Superconducting Cyclotron Laboratory, Michigan State University, East Lansing, Michigan 48824, USA}
\affiliation{Department of Physics $\&$ Astronomy, Michigan State University, East Lansing, Michigan 48824, USA}
\author{P.A.~DeYoung}
\affiliation{Department of Physics, Hope College, Holland, Michigan 49423, USA}
\author{J.E.~Finck}
\affiliation{Department of Physics, Central Michigan University, Mt. Pleasant, Michigan, 48859, USA}
\author{N.~Frank}
\affiliation{Department of Physics \& Astronomy, Augustana College, Rock Island, Illinois, 61201, USA}
\author{B.~Luther}
\affiliation{Department of Physics, Concordia College, Moorhead, Minnesota 56562, USA}
\author{E.~Lunderberg}
\altaffiliation[Present address: ]{National Superconducting Cyclotron Laboratory, Michigan State University, East Lansing, Michigan 48824, USA}
\affiliation{Department of Physics, Hope College, Holland, Michigan 49423, USA}
\author{M.~Jones}
\affiliation{National Superconducting Cyclotron Laboratory, Michigan State University, East Lansing, Michigan 48824, USA}
\affiliation{Department of Physics $\&$ Astronomy, Michigan State University, East Lansing, Michigan 48824, USA}
\author{S. Mosby}
\altaffiliation[Present address: ]{Los Alamos National Laboratory, Los Alamos, New Mexico 87545, USA}
\affiliation{National Superconducting Cyclotron Laboratory, Michigan State University, East Lansing, Michigan 48824, USA}
\affiliation{Department of Physics $\&$ Astronomy, Michigan State University, East Lansing, Michigan 48824, USA}
\author{J.~K.~Smith}
\altaffiliation[Present address: ]{TRIUMF, Vancouver, British Columbia V6T 2A3, Canada}
\affiliation{National Superconducting Cyclotron Laboratory, Michigan State University, East Lansing, Michigan 48824, USA}
\affiliation{Department of Physics $\&$ Astronomy, Michigan State University, East Lansing, Michigan 48824, USA}
\author{A. Spyrou}
\affiliation{National Superconducting Cyclotron Laboratory, Michigan State University, East Lansing, Michigan 48824, USA}
\affiliation{Department of Physics $\&$ Astronomy, Michigan State University, East Lansing, Michigan 48824, USA}
\author{M. Thoennessen}
\affiliation{National Superconducting Cyclotron Laboratory, Michigan State University, East Lansing, Michigan 48824, USA}
\affiliation{Department of Physics $\&$ Astronomy, Michigan State University, East Lansing, Michigan 48824, USA}

\date{\today}

\begin{abstract}
\begin{description}
\item[Background] Theoretical calculations have shown that the energy and angular correlations in the three-body decay of the two-neutron unbound $^{26}$O can provide information on the ground-state wavefunction, which has been predicted to have a dineutron configuration and $2n$ halo structure.
\item[Purpose]  To use the experimentally measured three-body correlations to gain insight into the properties of $^{26}$O, including the decay mechanism and ground-state resonance energy.
\item[Method]  $^{26}$O was produced in a one-proton knockout reaction from $^{27}$F and the $^{24}$O~$+ n + n$ decay products were measured using the MoNA-Sweeper setup.  The three-body correlations from the $^{26}$O ground-state resonance decay were extracted.  The experimental results were compared to Monte Carlo simulations in which the resonance energy and decay mechanism were varied.
\item[Results] The measured three-body correlations were well reproduced by the Monte Carlo simulations but were not sensitive to the decay mechanism due to the experimental resolutions. However, the three-body correlations were found to be sensitive to the resonance energy of $^{26}$O.  A $1\sigma$~upper-limit of 53~keV was extracted for the ground-state resonance energy of $^{26}$O.
\item[Conclusions]  Future attempts to measure the three-body correlations from the ground-state decay of $^{26}$O will be very challenging due to the need for a precise measurement of the $^{24}$O momentum at the reaction point in the target.
\end{description}
\end{abstract}



\maketitle

\par
\section{Introduction}
Exploring the limits of the chart of the nuclides provides fundamental benchmarks for theoretical calculations and opportunities to discover new phenomena~\cite{Tho04,Thoe11,Erler12,Bau07}.  Recently, experimental and theoretical studies of two-proton ($2p$) and two-neutron ($2n$) unbound systems have garnered significant interest~\cite{BAUMANN12,Aum14,Pfu12}.  These systems allow for the properties and decay mechanisms of nuclei with extreme neutron-to-proton ratios, existing beyond the driplines, to be examined.  Of particular interest are the correlations present in the decay of these three-body systems which can offer new insights into the initial structure and configuration of the nucleus~\cite{Pfu12,Mie07,Grig09,Ego12,Grig12,John10,Kik13,Aks13}. At the driplines, unique situations can present themselves in which the sequential decay process is forbidden and the ``true'' two-nucleon decay can be observed~\cite{Gold60}.  While substantial progress has been made in the theoretical descriptions of the $2p$ decay mechanism, resulting in impressive reproductions of experimental measurements, theoretical frameworks for describing the full three-body decay of $2n$ unbound systems are still at an early stage and appear to pose new challenges in comparison to the $2p$ decay~\cite{Pfu12,Grig09,Ego12,Grig12,Grigb_09,Gri13}.

\par
While the correlations from the decays of a wide range of $2n$ unbound systems have been measured ($^{5}$H~\cite{Gol05}, $^{10}$He~\cite{John10,Sid12}, $^{11}$Li~(excited state)~\cite{Smith14}, $^{13}$Li~\cite{John10,Kohley12}, $^{14}$Be~(excited state)~\cite{Aks13}, and $^{16}$Be~\cite{SPYROU12}), the $^{26}$O system provides a particularly interesting case to examine in light of recent experimental and theoretical work.  Currently, constraints on the $^{26}$O ground-state resonance energy of $< 200$~keV and $<40$~keV have been reported by the MoNA collaboration~\cite{LUN12} and R3B collaboration~\cite{Cae13}, respectively.  Thus, the sequential decay of $^{26}$O through $^{25}$O, with a ground state unbound by $\sim 770$~keV~\cite{Hof08,Cae13}, is forbidden and $^{26}$O must decay directly to $^{24}$O through the simultaneous emission of two neutrons.  These constraints have provided a sensitive observable for \emph{ab initio}-type calculations examining the role of three-nucleon forces and continuum effects~\cite{Hag12,Bog14,Herg13,Cip13}. Furthermore, theoretical calculations of Grigorenko \emph{et al.} suggested that $^{26}$O would be a candidate for $2n$ radioactivity assuming a pure $[d_{3/2}]^{2}$ neutron configuration and near threshold ground-state resonance~\cite{Gri11}. Through the use of the decay-in-target technique~\cite{Thoe13}, the MoNA collaboration extracted a half-life for $^{26}$O of 4.5$^{+1.1}_{-1.5}$(stat)~$\pm$~3(syst)~picoseconds which suggested the possibility for two-neutron radioactivity~\cite{Koh13}.  Based on the calculations of Grigorenko~\emph{et al.} a $t_{1/2}=4.5$~ps would only require the decay energy to be $<100$~keV~\cite{Gri11}.  However, the presence of even a small $[s_{1/2}]^{2}$ component in the $^{26}$O wavefunction would dominate the width of the decay (and lifetime).  New detailed calculations from Grigorenko \emph{et al.} using the three-body hyperspherical harmonics cluster model lowered the original constraint to $E_\textrm{decay} < 1$~keV for the measured $^{26}$O lifetime~\cite{Gri13}.  Estimates from the continuum shell-model, assuming a sequential decay, reported a similar constraint of $E_\textrm{decay} \lesssim 0.5$~keV~\cite{Volya14}.  Therefore, it is crucial that the $^{26}$O decay energy be further constrained by experimental measurements.

\par
In addition to the theoretical constraints on the $^{26}$O decay energy, full three-body calculations of the decay correlations have been reported by Grigorenko \emph{et al.}~\cite{Gri13} as well as by Hagino and Sagawa~\cite{Hag14}.  The results show that the three-body correlations  are sensitive to both the initial wavefunction and the properties of the decay mechanism including the final-state $nn$ interaction, recoil effects, and sub-barrier configuration mixing or rescattering of the $d-$wave neutrons into lower $\ell$ orbitals during the decay~\cite{Gri13,Hag14}.  The predictions of the ground state wavefunction added additional interest since the model of Grigorenko \emph{et al.} suggested $^{26}$O would have a strong halo structure with a rms radius of the valence neutrons around 5.7~fm~\cite{Gri13} and the model of Hagino and Sagawa showed the valence neutrons to be in a strong dineutron configuration~\cite{Hag14}.  Both theoretical frameworks predicted that these configurations would be manifested in the three-body decay correlations.  One specific signature shown in both models is that the distribution of angles between the two neutrons ($\theta_{nn}$) would be peaked near 180$^{\circ}$.

\par
In this article, we extract the experimental energy and angular correlations in the Jacobi coordinate system from the three-body decay of the $^{26}$O ground-state resonance. The experimental distributions are compared to a dineutron model, a phase space model, and the theoretical calculation from Hagino and Sagawa.  The sensitivity of the experimental results to the decay mode and ground-state resonance energy are presented.

\begin{figure*}
\includegraphics[width=0.7\textwidth]{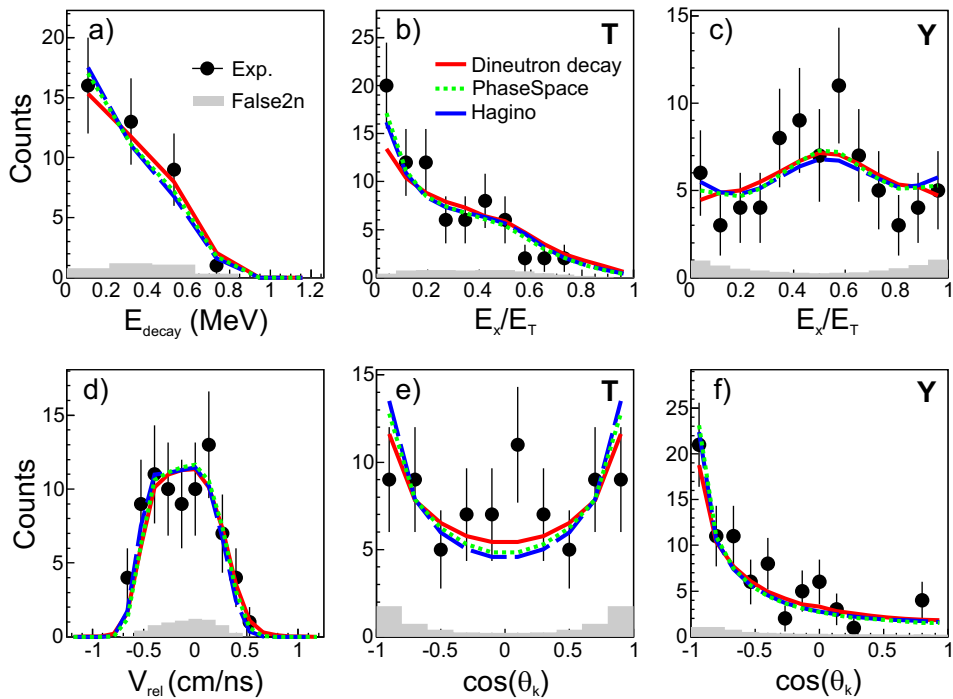}
\caption{\label{f:decay} (Color online) (a) Three-body decay energy spectrum, (b) Jacobi relative energy in \textbf{T} system, (c) Jacobi relative energy in the \textbf{Y} system, (d) relative velocity spectrum, (e) Jacobi angle in the \textbf{T} system, and (f) Jacobi angle in the \textbf{Y} system from the decay of $^{26}$O.  The experimental data are compared with simulations using three different decay modes and a half-life of 4~ps.  All results are gated on $E_\textrm{decay} < 0.7$~MeV and have the causality cuts applied.  The remaining false $2n$ component in the spectra, based on the \texttt{\sc Geant4} simulation, is shown as the solid grey area. }
\end{figure*}

\section{Experimental Details and Analysis}
\par
Since the experimental details have already been provided in Refs.~\cite{LUN12,Koh13,Koh14} where the ground state resonance and lifetime measurements were reported, only a brief overview is presented.  An 82~MeV/u $^{27}$F radioactive ion beam was produced at the National Superconducting Cyclotron Laboratory at Michigan State University from the projectile fragmentation of a 140 MeV/u $^{48}$Ca primary beam.  The $2n$-unbound $^{26}$O was produced from a one-proton knockout of the $^{27}$F secondary beam using a 705~mg/cm$^{2}$ Be reaction target.  The $^{26}$O decayed into $^{24}$O~+~$2n$ in the reaction target.  The triple coincidence measurement was accomplished using the MoNA-Sweeper setup~\cite{SWEEPER,MONA03,MONA05,INVTRACKING,Chr12}  and allowed for the invariant mass and correlations between the decay products of $^{26}$O to be determined.



\par
Measurements of $2n$ decays require the discrimination of the ``true'' $2n$ events from the ``false'' $2n$ background which is generated from a single neutron producing multiple hits within the array.  To remove the majority of false $2n$ events from the subsequent analysis, we applied the same causality cuts as used in our previous works reporting on $^{26}$O~\cite{LUN12,Koh13,Koh14}.  The causality cuts required the first two time ordered interactions within MoNA to have a spatial separation of $>$~25~cm and a relative velocity $>$~7~cm/ns.  After applying the causality cuts, the false $2n$ component present in the experimental spectra is nearly negligible.


\par
In order to isolate the ground state decay of $^{26}$O, only events with $E_\textrm{decay} < 0.7$~MeV were selected.  This selection was chosen to minimize any contribution from the decay of the first excited state of $^{26}$O. Recent calculations by Bogner \emph{et al.}~\cite{Bog14} and Hagino and Sagawa~\cite{Hag14b} indicate that the $2^{+}$ state is likely between 1 and 2~MeV.  The $E_\textrm{decay}$ spectrum with the 0.7~MeV cut is shown in Fig.~\ref{f:decay}(a).  The full $E_\textrm{decay}$ spectrum can be seen in Fig.~1 of Ref.~\cite{Koh13}.

\begin{figure}
\includegraphics[width=0.35\textwidth]{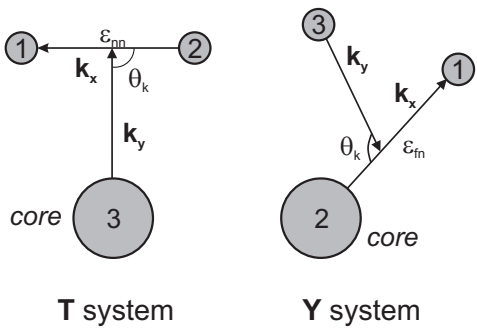}
\caption{\label{f:jacobi} Illustration of the \textbf{T} and \textbf{Y} Jacobi coordinate systems used to define the energy and angular correlations of the three body decay. }
\end{figure}

\par
The \textbf{T} and \textbf{Y} Jacobi coordinate systems, illustrated in Fig.~\ref{f:jacobi}, were used to define the angular and energy correlations in the three-body decay of $^{26}$O.  The three-body correlations can be fully described by the energy, $E_{x}$, and angle, $\cos(\theta_{k}$), defined in each Jacobi system as,
\begin{subequations}
\begin{align}
E_{x} &= \frac{(m_{1}+m_{2})k_{x}^{2}}{2m_{1}m_{2}} \\
\cos(\theta_{k}) &= \frac{\boldsymbol{k_{x}} \cdot \boldsymbol{k_{y}}}{k_{x}k_{y}}
\end{align}
\end{subequations}
with the Jacobi momenta $\boldsymbol{k_{x}}$ and $\boldsymbol{k_{y}}$ defined as,
\begin{subequations}
\begin{align}
\boldsymbol{k_{x}} &= \frac{m_{2}\boldsymbol{k_{1}} - m_{1}\boldsymbol{k_{2}}}{m_{1}+m_{2}} \\
\boldsymbol{k_{y}} &= \frac{m_{3}(\boldsymbol{k_{1}}+\boldsymbol{k_{2}}) - (m_{1}+m_{2})\boldsymbol{k_{3}}}{m_{1}+m_{2}+m_{3}}.
\end{align}
\end{subequations}
The mass and momentum of each particle is labeled as $m_{i}$ and $\boldsymbol{k_{i}}$, respectively.  As depicted in Fig.~\ref{f:jacobi}, $E_{x}$ represents the energy in the two-body system defined by particles 1 and 2, while $\theta_{k}$ represents the angle between that two-body system and particle 3 .  $E_{x}$ is often reported relative to the total three body decay energy ($E_{T}$). The experimental $E_{x}/E_{T}$ and $\cos(\theta_{k}$) distributions for the \textbf{T} and \textbf{Y} systems, with the causality cuts and $E_\textrm{decay} < $0.7~MeV criteria, are shown in panels (b), (c), (e), and (f) of Fig.~\ref{f:decay}.

\section{Simulations}
\par
Interpretation of the data requires comparisons with detailed simulations of the experimental setup.  A Monte Carlo simulation which included the incoming beam characteristics, reaction kinematics, and detector resolutions was utilized in the subsequent analysis~\cite{Koh12}.  The interactions of the neutrons within MoNA were modeled using the \texttt{\sc Geant4} framework~\cite{GEANT4,GEANT42} with the custom neutron interaction model~\textsc{menate\_r}~\cite{menateR,Koh12}.  This allowed the results of the Monte Carlo simulation to be treated identically to the experimental data.

\par
\begin{figure}
\includegraphics[width=0.39\textwidth]{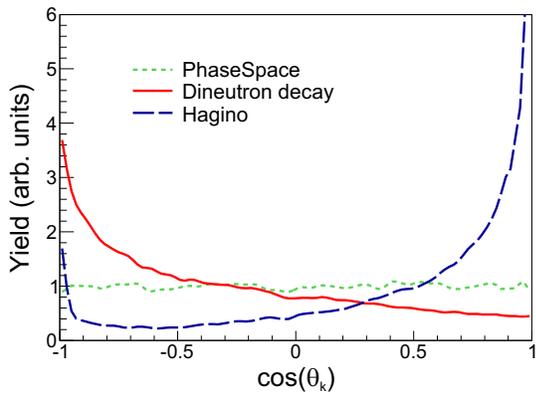}
\caption{\label{f:input} (Color online) Input $\cos(\theta_{k}$) distributions from the \textbf{Y} Jacobi coordinate system for the Monte Carlo simulations.  The distributions from the phase-space, dineutron, and Hagino decay models are presented.}
\end{figure}

The ground state resonance of $^{26}$O was simulated with a Breit-Wigner lineshape having a resonance energy, $E_{r}$, and a decay width, $\Gamma$.
The energy and angular correlations of the $^{24}$O~$+~n~+~n$ system  were simulated assuming three different decay models: a phase-space decay~\cite{3body,3bodyROOT}, a dineutron decay~\cite{Kohley12}, and a decay model based on the theoretical calculations of Hagino and Sagawa~\cite{Hag14}.  The phase-space decay model simulates the simultaneous breakup of $^{26}$O$ \rightarrow ^{24}$O$ + n +n$ assuming that the particles do not interact during the decay process~\cite{3body,3bodyROOT,SPYROU12,Kohley12}.  The dineutron decay model, described in Refs.~\cite{Kohley12,Volya06}, simulates a two-step process where a dineutron is emitted and then decays with an intrinsic energy defined by a $nn$-scattering length of $a_{s} =-18.7$~fm~\cite{Trot99}.  Lastly, a decay model was created based on the $\theta_{nn}$ angular distribution from the full three-body calculations of Hagino and Sagawa (Fig. 3 of Ref.~\cite{Hag14}).  Examples of the input $\cos(\theta_{k}$) distributions in the \textbf{Y} Jacobi system are shown in Fig.~\ref{f:input} for a 5~keV resonance energy.  As shown, the flat distribution is produced by the phase-space model indicating no correlations between the neutrons, whereas the dineutron and Hagino models produces distributions that are strongly peaked at $-1$ and 1 indicating the neutrons are emitted with opening angles near 0$^{\circ}$ and 180$^{\circ}$, respectively.  It is important to clarify that the dineutron decay model simulates the emission of a dineutron and does not necessitate the presence of a dineutron in the ground-state wavefunction of $^{26}$O.  In comparison, the correlations from Hagino and Sagawa are derived from a full three-body decay calculations which contain a dineutron configuration in the ground-state structure of $^{26}$O.

\par
The effects of a finite lifetime for $^{26}$O were also included in the Monte Carlo simulation.  As described in Ref.~\cite{Koh13}, a finite lifetime of $^{26}$O, in the range of picoseconds, will result in decreased velocity distributions of the two neutrons emitted after the $^{26}$O has traversed through part of the target.  The decreased neutron velocity is represented in the relative velocity spectrum, $V_\textrm{rel}$, defined as the difference in velocity between the neutron and the $^{24}$O fragment [Fig.~\ref{f:decay}(d)].


\section{Results and Discussion}

\par
The results from the simulations with the three different decay modes were simultaneously fit to the experimental decay energy, relative velocity, and Jacobi plots in Fig.~\ref{f:decay}.  The ground-state resonance energy was a free parameter varying from 0~keV to 100~keV in the fit. The width of the decay, $\Gamma$, was fixed at 1~keV since the width of the resonance is completely dominated by the experimental resolutions~\cite{LUN12,Koh13}. The influence of the $^{26}$O lifetime was also investigated with $t_{1/2}$~=~0~ps and 4~ps.  The best fit (shown in Fig.~\ref{f:decay}) of the phase-space, dineutron, and Hagino decay modes corresponds to $E_{r}$~=~15~keV, 15~keV, and 10~keV, respectively, with $t_{1/2}$~=~4~ps.

\par
The experimental results were well reproduced by all of the models indicating that the angular-energy correlations are relatively insensitive to the decay mode.  This is unexpected as previous works have shown $E_{x}/E_{T}$ from the \textbf{T} system and $\cos(\theta_{k}$) from the \textbf{Y} system to be particularly sensitive to the decay mode~\cite{SPYROU12,Kohley12,Koh14}.  For example, the input $\cos(\theta_{k}$) distribution should be strongly peaked at $-1$ for the dineutron decay, relatively flat for the phase-space decay, and peaked near 1 for the Hagino decay, as shown in Fig.~\ref{f:input}.  However, all simulations show the same correlation signatures when passed through the experimental filter.

\par
\begin{figure}
\includegraphics[width=0.31\textwidth]{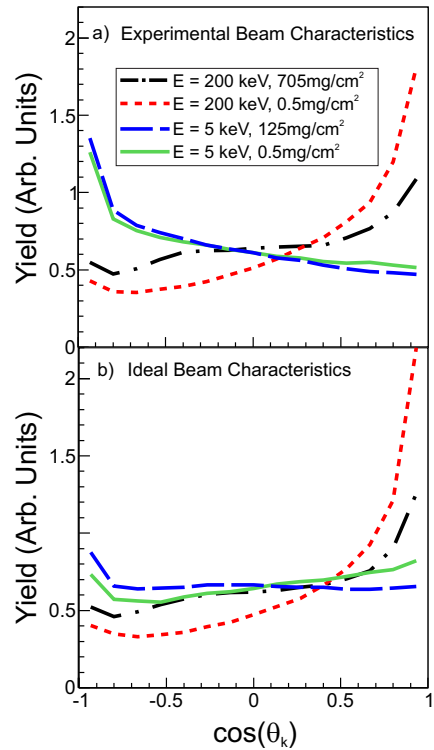}
\caption{\label{f:comp} (Color online) $\cos(\theta_{k})$ distributions from the \textbf{Y} Jacobi system for simulations using the Hagino decay mode with varying Be target thicknesses and decay energies for $^{26}$O. The results are shown (a) with and (b) without the inclusion of the experimental incoming beam characteristics in the simulations.}
\end{figure}

The insensitivity of the results to the decay mode is due to the uncertainty in the momentum of the $^{24}$O at the reaction point in the target and the near threshold energy of the $^{26}$O  ground-state resonance.  The Jacobi variables, which describe the three-body correlations, are directly related to the relative momenta of the particles in the decay.  When the decay energy is very low, the relative momentum between the particles is then very small.  Thus, a small uncertainty in the momentum of the $^{24}$O fragment can cause a large change in the relative momentum between the neutrons and fragment.  This produces a false enhancement in the correlation between the two neutrons relative to the $^{24}$O fragment which is manifested as the observed dineutron signatures shown in panels (b) and (f) of Fig.~\ref{f:decay}.  This effect is diminished with increasing decay energy.  For example, in Fig.~\ref{f:comp}(a) the the $\cos(\theta_{k}$) distribution is shown for a resonance energy of 200~keV with the Hagino decay model (black dash-dot line).  With the 200~keV resonance energy, the angular distribution peaks at $\cos(\theta_{k}$)~=~$1$ reproducing the overall feature of the input distribution from Hagino.

\par
The uncertainty in the momentum of the $^{24}$O stems mainly from two factors: (1) the uncertainty in the position at which the $^{27}$F($-p$) reaction occurs within the target and (2) the size and angular spread of the incoming beam.  While the momentum, or $B\rho$, of the outgoing $^{24}$O is defined by the path of the fragment through the Sweeper magnet, the momentum at the reaction point requires the location of the reaction in the target to be known so that the energy-loss within the target can be accurately calculated.  Since the reaction location is unknown, it was assumed that the reaction took place at the mid-point of the target in the present analysis.  In Fig.~\ref{f:comp}(a), the effect of the target thickness is shown.  For a 200~keV resonance, decreasing the target thickness from 705~mg/cm$^{2}$ to 0.5~mg/cm$^{2}$ improves the agreement between the results of the simulation and the input Hagino distribution.  However, decreasing the target thickness does not improve the results when $E_{r}$~=~5~keV and the dineutron signature still prevails.  This is due to the characteristics of the incoming $^{27}$F beam which can produce uncertainties in the $B\rho$ analysis of the $^{24}$O through the Sweeper magnet.  Again, if the decay energy of $^{26}$O is near threshold any uncertainty in the $^{24}$O momentum is magnified in the three-body correlations.  From the experiment, the beam spot on the target was estimated to be 5~mm x 5~mm and the incoming angular distribution of the beam was approximated as a Gaussian distribution with a mean of 0$^{\circ}$ and $\sigma$~=~7~mrad (2~mrad) in the dispersive (non-dispersive) direction.  These deviations from an ideal beam could be corrected event-by-event using tracking detectors upstream of the target, however such detectors were not available for the present experiment.

\par
Fig.~\ref{f:comp}(b) shows the expected angular distributions assuming ideal beam characteristics.  While the simulations with the 200~keV resonance energy are not significantly altered with the ideal beam characteristics, the dineutron signature is no longer present in the simulation with a 5~keV resonance energy.  A relatively flat $\cos(\theta_{k}$) distribution is observed with a 125~mg/cm$^{2}$ target and a slight peaking towards $\cos(\theta_{k}$)~=~1 is shown with a 0.5~mg/cm$^{2}$ target with $E_{r}$~=~5~keV.  The results from the simulation, therefore, indicate that an experimental measurement sensitive to the angular and energy correlations in the three body decay of $^{26}$O will be extremely difficult assuming the ground state resonance is near threshold. Even with an increased $^{27}$F beam rate, the use of a 0.5~mg/cm$^{2}$ target would pose a significant challenge in obtaining the required statistics to identify the decay mode of $^{26}$O.  While future experiments will likely encounter a similar situation, it is important to emphasis that these results and sensitivity studies are specific to the present experiment and MoNA-Sweeper setup.

\section{Constraints on the $^{26}$O ground-state resonance energy}

\par
Although not sensitive to the decay mode, the measured three-body correlations are sensitive to the $^{26}$O decay energy and, therefore, can allow for improved constraints to be extracted through a self-consistent fit of all the spectra presented in Fig.~\ref{f:decay}.  As mentioned previously, the width of the resonance was kept constant, $\Gamma$~=~1~keV.  The fitting procedure was completed for $t_{1/2} = 0$~ps and 4~ps to examine the half-life dependence of the results.  Since the $V_\textrm{rel}$ distribution [Fig.~\ref{f:decay}(d)] has been shown to be best fit with $t_{1/2} \sim 4$~ps~\cite{Koh13}, it was not included in the fitting procedure with $t_{1/2}=0$~ps.

\begin{figure}
\includegraphics[width=0.49\textwidth]{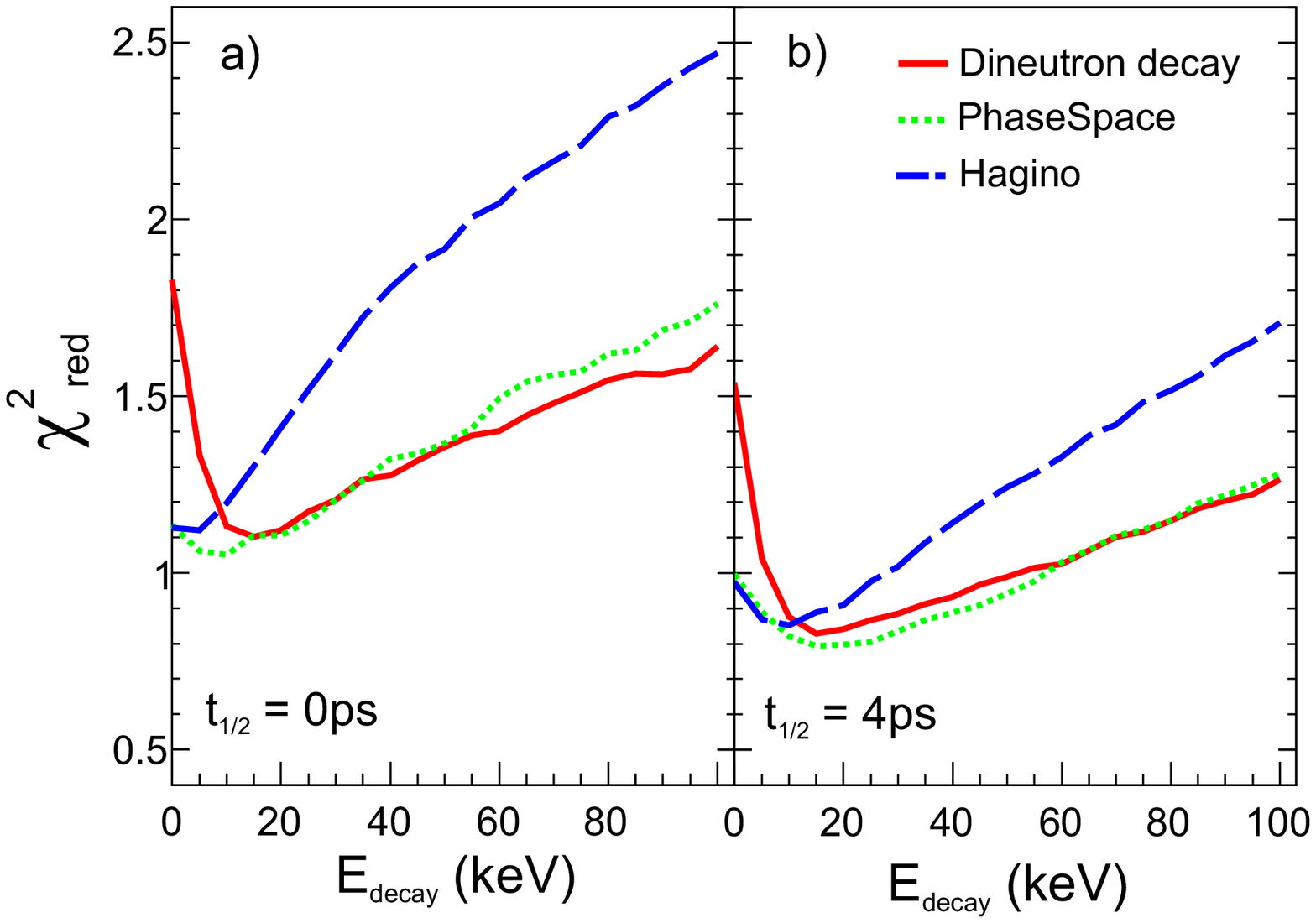}
\caption{\label{f:chi} (Color online) Reduced chi-squared as a function of $^{26}$O ground-state decay energy for (a) $t_{1/2}$~=~0~ps and (b) 4~ps with the three different decay modes used in the simulation.}
\end{figure}

\begin{table}
\begin{center}
\caption{$1\sigma$ limits on the $^{26}$O ground-state resonance energy extracted from the simultaneous fitting of the three-body correlations and decay energy spectrum with different decay modes in the simulation.}
\begin{tabular}{c c c}
\hline
\hline
   &\multicolumn{2}{c}{1$\sigma$ limit}\\
Model  &$t_{1/2} = 0$~ps   &$t_{1/2} = 4$~ps\\
\hline
Hagino &$< 15$~keV  &$< 31$~keV\\
Phase-Space &$< 33$~keV  &$< 53$~keV\\
Dineutron &$6 - 42$~keV  &$6 - 53$~keV\\
\hline
\hline
\end{tabular}
\label{table}
\end{center}
\end{table}

\par
The results of the simultaneous fitting procedure for $t_{1/2}=0$~ps and 4~ps are presented in Fig.~\ref{f:chi} where the reduced chi-squared value ($\chi^{2}_{red}$) from each fit is shown as a function of the $^{26}$O resonance energy for the three simulated decay modes.  In all cases, it is clear that the inclusion of the three-body correlations into the fit greatly enhances the sensitivity of the results to the decay energy relative to our previous work where a constraint of $E_\textrm{decay} < 200$~keV was extracted using only the decay energy spectrum~\cite{LUN12}.  The strong correlations present in the input distributions of the Hagino and dineutron decay models (Fig.~\ref{f:input}) produce a strong sensitivity of the results at high and low decay energies, respectively.  At low decay energies, the correlation between the emitted neutrons are severely overestimated by the dineutron decay resulting in the large $\chi^{2}_{red}$.  Similarly, as the decay energy increases the correlations from the Hagino decay, with $\cos(\theta_{k})$ peaking near 1, become more prominent producing a rapid increase in $\chi^{2}_{red}$.  The phase-space decay is the least sensitive to the decay energy since it does not exhibit such strong features in the correlations.

\par
For all three models, the $\chi^{2}_{red}$ rises with the decay energy increasing beyond $\sim$20~keV due to the poor fitting of the three-body correlations.  From the chi-squared fits, the $1\sigma$ limits for the ground-state $^{26}$O resonance energy were extracted and are shown in Table~\ref{table}.  The results strongly support the conclusion that the ground-state resonance energy is near threshold, with upper-limits ranging from 15~keV to 50~keV depending on the decay mode and half-life.   The extracted $1\sigma$ limits are in excellent agreement with the 40~keV upper-limit reported by the R3B collaboration~\cite{Cae13}.  It is worth noting that the results using the correlations from the full three-body calculations of Hagino and Sagawa~\cite{Hag14} (which also agree with the detailed calculations of Grigorenko \emph{et al.}~\cite{Gri13}) should provide the most realistic model for the decay and produces $1\sigma$ upper-limits of 15~keV and 31~keV for the cases of $t_{1/2}=0$~ps and 4~ps, respectively.

\section{Conclusions}
The three-body energy and angular correlations in the ground-state resonance decay of $^{26}$O$\rightarrow ^{24}$O$~+~n~+~n$ were experimentally measured using the MoNA-Sweeper setup.  The experimental results were compared to Monte Carlo simulations using three different decay models: a phase-space model, dineutron model, and three-body decay model based on the theoretical calculations of Hagino and Sagawa~\cite{Hag14}.  The experimental three-body correlations were well reproduced by all three of the decay models indicating an insensitivity of the experimental data to the decay mode. This was shown to be due to the low decay energy of $^{26}$O which, therefore, requires very precise measurements of the relative momentum of the $^{24}$O and two neutrons to reconstruct the correlations.  Monte Carlo simulations showed that the target thickness and tracking of the incoming beam characteristics largely define the experimental resolutions for measuring the three-body correlations.  Even with improving these aspects of the experiment, the simulation results indicate that future attempts to measure the three-body correlations from the decay of $^{26}$O will face a difficult challenge due to the near threshold ground-state resonance energy.

\par
Through simultaneous and self-consistent fitting of the decay energy spectrum and the Jacobi three-body correlation variables strict constraints on the $^{26}$O decay energy were extracted.  While the results were dependent on the decay mode and half-life of $^{26}$O, a maximum 1$\sigma$ upper-limit of 53~keV was obtained.  Furthermore, if one assumes that the correlations from the full three-body decay calculations of Hagino and Sagawa~\cite{Hag14} are correct (noting that similar correlations are also predicted by Grigorenko \emph{et al.}~\cite{Gri13}), then the $1\sigma$ upper-limit for the $^{26}$O ground-state is 31~keV.  Additional measurements, requiring increased statistics, are strongly desired to provide a precise measurement ground-state resonance energy for $^{26}$O.

\begin{acknowledgments}   We would like to thank Anthony Kuchera (NSCL) and Paul Gueye (Hampton University) for proof reading the
manuscript.  The authors gratefully acknowledge the support of the NSCL operations staff for providing a high quality beam. This work is supported by the National Science Foundation under Grant No. PHY-1102511 and PHY-1306074.
\end{acknowledgments}


%

\end{document}